\documentclass[sigconf]{acmart}
\AtBeginDocument{%
  }

\usepackage{multirow}
\usepackage{enumitem}
\usepackage{hyperref}
\usepackage{makecell}
\usepackage{booktabs}
\usepackage{siunitx}
\usepackage{array}
\usepackage[normalem]{ulem}

\copyrightyear{2026}
\acmYear{2026}
\setcopyright{cc}
\setcctype{by}
\acmConference[CIKM '26] {Proceedings of the 35th ACM International Conference on Information and Knowledge Management}{November 7--11, 2026}{Rome, Italy.}
\acmBooktitle{Proceedings of the 35th ACM International Conference on Information and Knowledge Management (CIKM '26), November 7--11, 2026, Rome, Italy}
\acmISBN{979-8-4007-2539-5/2026/11}
\acmDOI{10.1145/3799682.3840604}
\begin{document}

\title[A Multi-Target Framework for Learning Token-Weighted Objectives in Generative Recommenders]{Beyond Uniform Token Training: A Multi-Target Framework for Learning Token-Weighted Objectives in Generative Recommenders}


\author{Wei-Ning Chiu}
\email{wchiu2@andrew.cmu.edu}
\orcid{1234-5678-9012}
\affiliation{%
  \institution{Carnegie Mellon University}
  \city{Pittsburgh}
  \country{United States}
}

\author{Song-Duo Ma}
\email{r14944001@ntu.edu.tw}
\orcid{1234-5678-9012}
\affiliation{%
  \institution{National Taiwan University}
  \city{Taipei}
  \country{Taiwan}
}

\author{Han-Jay Shu}
\email{Jerryshu98@gapp.nthu.edu.tw}
\orcid{1234-5678-9012}
\affiliation{%
  \institution{National Tsing Hua University}
  \city{Hsinchu}
  \country{Taiwan}
}

\author{Chuan-Ju Wang}
\email{cjwang@citi.sinica.edu.tw}
\affiliation{%
  \institution{Academia Sinica}
  \city{Taipei}
  \country{Taiwan}
}
\author{Pu-Jen Cheng}
\email{pjcheng@csie.ntu.edu.tw}
\affiliation{%
  \institution{National Taiwan University}
  \city{Taipei}
  \country{Taiwan}
}

\renewcommand{\shortauthors}{Chiu et al.}

\begin{abstract}
  Recent generative recommendation models recast next-item prediction as the generation of a semantic identifier sequence. While this formulation enables autoregressive models to produce item IDs directly, the commonly used token-level likelihood objective does not distinguish between tokens that play different roles in item identification. This limitation is especially pronounced for semantic-ID representations, where prefix tokens often determine coarse item groups and later tokens provide finer-grained disambiguation. To better align training with the structure of semantic IDs, we study token-level learning signals from two complementary perspectives. First, we introduce a prefix-aware weighting scheme, Front-Greater Weighting, which emphasizes tokens according to their contribution to reducing semantic ambiguity among candidate items. Second, frequency weighting increases the learning emphasis on infrequent tokens, addressing the long-tailed distributions and popularity bias commonly observed in recommendation data. We further introduce a multi-target optimization framework with curriculum learning, which integrates the two token-weighted objectives with the standard likelihood and enables stable optimization with adaptive emphasis across training stages. Experiments on multiple benchmark datasets demonstrate that the proposed approach consistently improves generative recommendation performance over strong baselines and prior token-weighting methods. Additional analyses show that the method is robust across different semantic-ID constructions and backbone scales, and that it improves recommendation quality for both popular and long-tail items.
  Code is available at 
  \href{https://github.com/CHIUWEINING/Token-Weighted-Multi-Target-Learning-for-Generative-Recommenders-with-Curriculum-Learning}{\texttt{\underline{github repository}}}\footnote{\url{https://github.com/CHIUWEINING/Token-Weighted-Multi-Target-Learning-for-Generative-Recommenders-with-Curriculum-Learning}}.
\end{abstract}




\maketitle
\section{Introduction}
With the rapid advancement of large language models (LLMs), a growing body of research has explored how these models can be incorporated into recommendation pipelines \cite{p5, bigrec, llm4rec_survey, huang2025augment}.  Among many paradigms \cite{wei2024llmrec, sun2025llmser, trsr, llm4isr}, generative recommendation directly treats the language model itself as the recommender. 
In this formulation, next-item prediction is cast as a sequence-to-sequence task, similar to generating text in natural language QA settings \cite{tiger, genrec}.
Recent progress has explored a wide range of item representations for generative recommendation, from raw item titles to learned semantic identifiers such as those produced by Residual Quantization (RQ) \cite{rqvae}. These codebook-based IDs have demonstrated strong advantages in capturing hierarchical semantics, integrating collaborative information~\cite{letter}, and remaining compatible with autoregressive decoding architectures \cite{lcrec, letter}.

Despite these advances, we identify a fundamental misalignment in how current generative recommendation models are trained. Most existing approaches rely on a standard next-token negative log-likelihood (NLL) and implicitly treat all tokens as equally important \cite{p5, genrec, bigrec}. However, in generative recommendation, tokens differ substantially in their contribution to identifying the correct item \cite{igd}. This issue persists even in codebook-based representations.
\begin{figure}[t]
\centering
\includegraphics[width=0.95\linewidth]{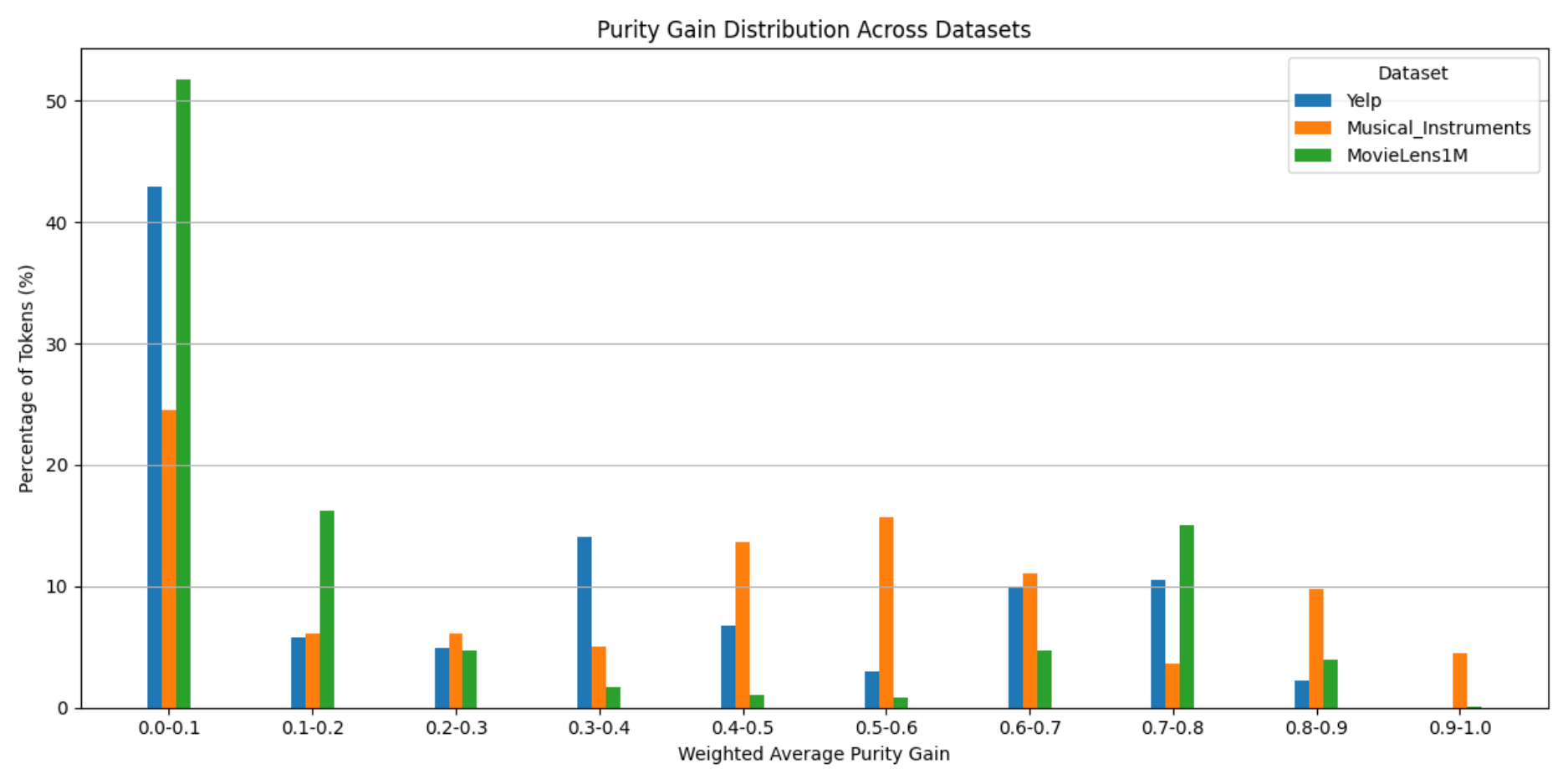}
\caption{Average purity gain per token. For each token, purity gain is defined as the increase in prefix purity when the token is appended to a prefix. Purity is computed as
$1 - \frac{H}{H_{\max}}$,
where $H$ denotes the entropy of item frequencies under the prefix and $H_{\max} = \log(n)$, with $n$ being the number of items sharing the prefix. The reported value is the average purity gain of each token across all its occurrences. Most tokens contribute little purity gain, while a small subset plays a dominant role in identifying items.}
\label{fig:purity_gain}
\end{figure}


\begin{table}[t]
\centering
\small
\setlength{\tabcolsep}{5pt}
\begin{tabular}{lcccc}
\toprule
Dataset & Layer-1 & Layer-2 & Layer-3 & Layer-4 \\
\midrule
Musical Instruments~\cite{amazon23}
& 99.56\% & 88.03\% & 62.50\% & 62.47\% \\

Industrial and Scientific~\cite{amazon23}
& 99.60\% & 97.64\% & 85.17\% & 35.05\%\\

Yelp\footnote{Yelp dataset available at \url{https://business.yelp.com/data/resources/open-dataset/}}
& 99.60\% & 96.14\% & 44.01\% & 12.13\% \\

MovieLens 1M~\cite{movielens}
& 99.60\% & 92.39\% & 14.29\% & 1.96\% \\ 
\bottomrule
\end{tabular}
\caption{Average ratio of items filtered out per layer when an additional token is added. We use four codebooks
with sizes [256, 256, 256, 256] for our RQ-VAE. }
\label{tab:layer-imp}
\end{table}

To illustrate this, we generate RQ-based semantic IDs with RQ-VAE~\cite{rqvae} and report the average purity gain of each token, which is the increase in normalized purity when each token is appended to a prefix, in Figure~\ref{fig:purity_gain}. The results show that tokens differ substantially in their contributions to purity. Moreover, consistent with observations in Lin et al.~\cite{igd}, most tokens exhibit very low purity gain, whereas a few key positions provide disproportionately large purity gains.
We further examine the filtering power of each token position by measuring the average fraction of items filtered out when extending the semantic-ID prefix by one token. As shown in Table~\ref{tab:layer-imp}, early-position tokens prune a substantially larger proportion of the remaining candidate items, highlighting their greater discriminative power.

\begin{figure}[t]
\centering
\includegraphics[width=0.95\linewidth]{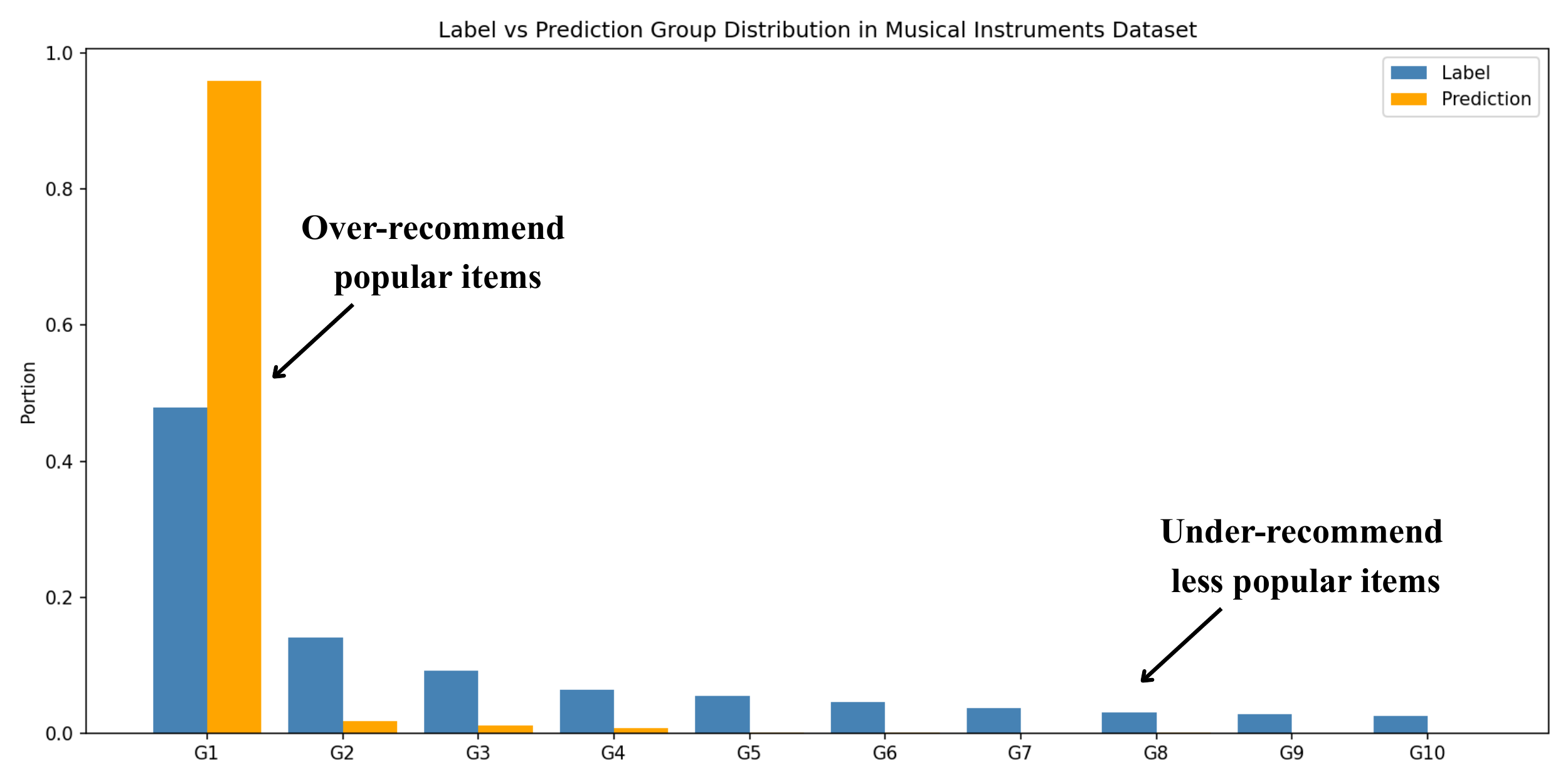}
\caption{Comparison between the empirical item-occurrence distribution (ground-truth labels) and TIGER’s predicted item distribution on the \textit{Musical Instruments} (Amazon'23) dataset~\cite{amazon23}. We split item frequency into ten quantiles (deciles) and count how many items fall into each quantile. TIGER exhibits a strong tendency to over-recommend popular items while under-representing rare items, reflecting the inherent bias in generative recommenders under long-tailed distributions.}
\label{fig:overrecommend}
\end{figure}
In addition to positional asymmetry, recommendation data exhibit long-tailed item and token distributions. Although such imbalances arise naturally from user behavior, in Figure~\ref{fig:overrecommend}, we show that TIGER~\cite{tiger}, a representative generative recommender based on RQ semantic IDs, can suffer from popularity bias, where it disproportionately recommends popular items while overlooking less frequent ones. These observations motivate the need for token-weighting strategies that explicitly address frequency-induced bias.
To address these issues, we introduce two token-weighting strategies that model information gain from perspectives intrinsic to codebook-based semantic IDs, rather than relying on Shannon entropy over generic token distributions (details in Section~\ref{sec:token_weight}):

\begin{itemize}
    \item \textbf{Front-Greater Weighting.}  
    Leveraging the hierarchical structure of RQ-based IDs and autoregressive decoding, we quantify \emph{semantic conditional information gain} by measuring how much additional semantic uncertainty is reduced when a token is revealed given its prefix. Tokens that yield larger reductions in semantic dispersion (i.e., typically early-position tokens) are assigned higher weights.

    \item \textbf{Frequency Weighting.}  
    To mitigate popularity bias induced by long-tailed item and token distributions, we model \emph{marginal information gain under the empirical data distribution} using the effective number of samples~\cite{effectiveNum}. Tokens with lower frequency, whose occurrences provide stronger learning signals, receive higher weights.
\end{itemize}

Beyond designing token-weighting strategies, we introduce a \textbf{multi-target framework with curriculum learning} (detailed in Section~\ref{sec:curr-aware weighting scheme}). We view each weighting strategy, namely Front-Greater, Frequency, and standard NLL (i.e., original cross-entropy likelihood), as a separate objective. For each objective, we introduce a learnable parameter that is normalized and applied to adjust the softmax logits, thereby controlling the strength of the associated gradient signal during training. Through theoretical derivation, we show that these normalized logit-scaling parameters admit an equivalent interpretation as objective weights in the resulting training objective. Building on this interpretation, we apply curriculum learning over the derived weights, which emphasizes Front-Greater and standard NLL early in training (promoting stability), and gradually shifts toward Frequency weighting later on (promoting fine-grained discrimination). This paradigm enables the model to adaptively determine the importance of each objective over time.

We evaluate our method across a wide range of experimental settings. Overall, it consistently outperforms existing token-weighting methods, achieving average gains of 6.11\% (Hit@5), 7.14\% (NDCG@5), 3.39\% (Hit@10), and 5.52\% (NDCG@10) over the base TIGER model. Ablation studies confirm the necessity of each component. Additional experiments demonstrate generalizability across semantic-ID types and model scales, improved robustness, and substantial superiority on both head and tail items.

\textbf{Our contributions are summarized as follows:}
\begin{itemize}
    \item We identify fundamental challenges in generative recommendation with semantic IDs and propose two complementary information-gain–based token-weighting strategies, namely \textbf{Front-Greater} and \textbf{Frequency} weighting, to explicitly address them.
    \item We introduce a \textbf{multi-target framework with curriculum learning} that jointly optimizes multiple token-weighted objectives with standard likelihood, enabling principled integration and stable optimization across different learning stages.
    \item We conduct extensive experiments across multiple datasets, semantic-ID constructions, model sizes, and evaluation settings, demonstrating consistent improvements over strong baselines and highlighting the robustness and general applicability of our approach for generative recommendation.
\end{itemize}

\section{Related Work}
We briefly review prior studies on LLM-based recommender systems, followed by works on token-level modeling for generative recommenders.
\subsection{LLM-based Recommender Systems}

Early LLM-based recommenders adapt pre-trained text-to-text models by converting user–item interactions into natural language sequences and applying prompting or fine-tuning. P5~\cite{p5} unifies multiple recommendation tasks under a pretrain–prompt–predict framework using a T5-style language modeling objective. Subsequent works improve efficiency and interpretability: POD~\cite{pod} distills discrete prompts into continuous prompt embeddings, while RDRec~\cite{rdrec} distills LLM-generated rationales from reviews to explicitly model user preferences. Instruction-tuned generative recommenders further tailor LLMs to recommendation tasks. TallRec~\cite{tallrec} formulates recommendation as a binary decision conditioned on user history, whereas GenRec~\cite{genrec} directly generates target items autoregressively from item text. BIGRec~\cite{bigrec} introduces an additional grounding stage to align generated tokens with valid item identifiers, improving consistency with ranking metrics.

Another line of work represents items using learned semantic identifiers (SIDs) based on vector or residual quantization~\cite{pq,pqcf,vqpq,rqkmeans,rvq}. Building on this idea, TIGER~\cite{tiger} generates hierarchical semantic IDs with RQ-VAE~\cite{rqvae} and trains a generative model to predict them in a coarse-to-fine manner, improving robustness under cold-start settings. LETTER~\cite{letter} further improves ID quality by jointly optimizing semantic structure, collaborative alignment, and codebook diversity, while LC-Rec~\cite{lcrec} integrates collaborative signals directly into learned identifiers through multi-task training. Beyond symbolic IDs, GRAM~\cite{gram} maps hierarchical semantic clusters to existing LLM vocabulary tokens, preserving structure while better leveraging the LLM’s linguistic prior. CoFiRec~\cite{cofirec} hierarchically constructs item identifiers using multiple vector quantization (VQ) modules, rather than representing each item with a single embedding.  

Despite these advances, most existing methods treat all tokens equally during training and overlook token-level optimization, which is a limitation our work explicitly addresses.
\subsection{Token-Weighting on Generative Recommenders}
Although most generative recommenders optimize a standard next-token cross-entropy loss that treats all tokens equally, recent studies challenge this assumption by explicitly modeling token-level contributions during training or decoding.
~\citet{letter} propose ranking-guided loss, which incorporates the temperature parameter into the logits to better handle hard-negatives. Counterfactual Fine-Tuning (CFT)~\cite{cft} explicitly emphasizes the role of behavior sequences with a token-weighting strategy that upweights earlier tokens.
IGD~\cite{igd} formulates item generation as a decision process and measures token importance via information gain computed from Shannon entropy~\cite{shannon}. 
While prior token-level methods often employ heuristic positional weighting or importance measures derived from generic token distributions (e.g., Shannon entropy), our method is fundamentally grounded in the unique properties of codebook-based semantic IDs, whose hierarchical, prefix-decomposable structure allows us to quantify token contribution such as prefix-conditional uncertainty reduction. 

\section{Preliminary}
Before introducing our proposed method, we briefly describe the core components used throughout this work. We first explain how codebook-based item identifiers (IDs) are constructed using Residual Quantization VAE (RQ-VAE)~\cite{rqvae}, and then show how these IDs enable recommendation to be formulated as a sequence-to-sequence generation task. 

\subsection{Codebook-Based ID Generation via Residual Quantization VAE}

With the rise of generative retrieval~\cite{dsi} and generative recommendation, RQ-VAE has become a well-known approach for constructing semantic item identifiers due to its robustness and natural compatibility with language-model decoding~\cite{tiger,lcrec}. Given item metadata such as titles or descriptions, RQ-VAE maps each item to a fixed-length sequence of discrete codes:
\[
\text{Item Metadata} \;\rightarrow\; [c_1, c_2, \dots, c_L],
\]
where $L$ denotes the number of codebooks and each code $c_i$ corresponds to an index in a learned codebook.

RQ-VAE performs residual quantization in multiple stages, where each codebook encodes the residual information left by previous stages. This process yields a hierarchical, coarse-to-fine semantic representation. As a result, each item is represented by a structured sequence of tokens that is well suited for autoregressive generation.

\subsection{Generative Recommendation with Codebook-Based IDs}

We now describe how these codebook-based IDs are used in generative recommendation. In sequential recommendation, a user’s interaction history
\[
h = (\mathrm{item}_1, \mathrm{item}_2, \dots, \mathrm{item}_{n-1})
\]
is converted into a sequence of semantic ID tokens by replacing each item with its $L$-token code, and the entire input interaction sequence is thus flattened into
\[
x = [c^1_{1}, c^1_{2}, \dots, c^1_{L},\;
      c^2_{1}, \dots, c^2_{L},\; \dots,\;
      c^{n-1}_{1}, \dots, c^{n-1}_{L}],
\]
and the target output is the ID of the next item:
\[
y = [c^n_{1}, c^n_{2}, \dots, c^n_{L}].
\]

A generative recommender then autoregressively predicts the target ID sequence, treating recommendation as a sequence-to-sequence generation task. During inference, constrained beam search is used to generate a ranked list of valid item IDs. 

\section{Methodology}
In this section, we present the core components of our method. We begin by introducing two token-level weighting strategies from an information-gain perspective, both designed to improve the training dynamics of generative recommenders. We then describe our multi-target framework with curriculum learning for mixing multiple loss objectives throughout training.

\subsection{Token Weighting Strategies}
\label{sec:token_weight}
In generative language models, training is typically performed using the token-level cross-entropy loss. 
However, in codebook-based semantic item representations, tokens are not equally informative. To account for this asymmetry, we incorporate token-level weighting into the loss function. Specifically, we assign each token position $i$ a weight $w_i$, modifying the loss to
\[
\mathcal{L}_{\text{weighted-CE}} = - \sum_{i=1}^{L} w_i \, \log P_{\theta}(y_i \mid y_{<i}, x),
\]
where $P_{\theta}(y_i \mid y_{<i}, x)$ is the model's predicted probability of token $y_i$ conditioned on the input $x$ and previously generated tokens $y_{<i}$.
This generalized formulation enables us to emphasize tokens that contribute more semantic information or require additional learning emphasis.

Building upon this framework, we propose two complementary token-weighting strategies grounded in the structural and distributional properties of codebook-based semantic IDs. Front-Greater Token Weighting captures semantic conditional information gain induced by hierarchical prefixes, while Frequency Token Weighting models marginal information gain under long-tailed token distributions. Unlike Shannon entropy–based weighting~\cite{shannon}, both are tailored specifically to generative recommendation with semantic IDs.

\subsubsection{\textbf{Front-Greater Token Weighting}}
In sequential recommendation, a prediction is considered correct only when \emph{all} tokens match the target sequence exactly. In particular, an error in any early token invalidates the entire prediction even if later tokens are correct. Despite this asymmetric influence, standard cross-entropy assigns equal weight to all token positions during training. To address this mismatch, we introduce \textbf{Front-Greater Token Weighting}, which assigns larger weights to earlier tokens. The key is to quantify how much semantic disambiguation each token position provides.

\paragraph{Quantifying semantic gain via prefix-based embedding concentration.}
We leverage the hierarchical structure of semantic IDs to quantify semantic information gain, i.e., how much revealing a token reduces ambiguity in the underlying item space. 
For each prefix length $k$, we partition items into groups $\mathcal{G}_k$ based on their shared top-$k$ tokens, i.e., their length-k ID prefixes. For any group $G\in\mathcal{G}_k$, let $\{u_i\}_{i\in G}\subset\mathbb{R}^d$ denote the corresponding item embeddings and define the group centroid
\[
\bar{u}_G \;=\; \frac{1}{|G|}\sum_{i\in G} u_i.
\]
We measure within-group concentration by the mean squared distance to the centroid:
\[
V(G) \;=\; \frac{1}{|G|}\sum_{i\in G}\|u_i-\bar{u}_G\|_2^2.
\]
Let $I$ be a uniformly sampled item index from $\{1,\dots,N\}$ and let $G_k(I)$ denote the group containing $I$ under prefix length $k$. We define the expected dispersion at level $k$ as
\[
\mu_k \;:=\; \mathbb{E}_{I}\!\left[\,V\!\left(G_k(I)\right)\right].
\]
As longer prefixes induce progressively finer partitions of the item space, $\mu_k$ captures the residual semantic uncertainty of an item \emph{conditioned on its top-$k$ prefix}. We therefore define the semantic disambiguation contributed by the $k$-th token as the reduction in expected dispersion:
\[
\delta_k \;:=\; \mu_{k-1}-\mu_k,
\qquad
w_k^{\mathrm{fg}} \propto \max(\delta_k,0).
\]
In this sense, $\delta_k$ serves as a geometric surrogate of the \emph{conditional information gain} contributed by token $k$ under prefix-based refinement.
Intuitively, a token receives a larger weight if revealing it leads to a greater expected reduction in within-group spread, indicating stronger conditional semantic disambiguation.

The following lemma formalizes that refining ID prefixes cannot increase the expected Fr\'echet variance~\cite{frechet1948}, which measures within-group dispersion as the mean squared distance to the group centroid~\cite{dubey2019frechet}. Therefore, $\delta_k$ is non-negative in expectation.
\begin{lemma}[Prefix Tokens Reduce Semantic Dispersion]
\label{lemma:dispersion}
Assume items $\mathcal{I}$ are embedded in a Euclidean space $\mathbb{R}^d$ and are hierarchically grouped by the prefix of length $k$ of their semantic ID sequences (so the partition at level $k$ refines that at level $k-1$). Let $\mathbb{E}[V_k]$ denote the expected Fr\'echet variance (i.e., the average within-group dispersion) over all items at prefix level $k$, i.e., $\mathbb{E}[V_k]=\mathbb{E}_I\!\left[V\!\left(G_k(I)\right)\right]$, where $G_k(I)$ denotes the group containing items $I$ at prefix level $k$ and $V(G)$ measures within-group dispersion as the average squared distance to the group centroid. Then $\mathbb{E}[V_k]$ is non-increasing with respect to $k$:
\[
\mathbb{E}[V_k] \;\le\; \mathbb{E}[V_{k-1}],
\]
and the reduction $\delta_k=\mathbb{E}[V_{k-1}] - \mathbb{E}[V_k]$ quantifies the conditional semantic information gain contributed by the $k$-th token.
\end{lemma}
Lemma~\ref{lemma:dispersion} follows directly from the fact that refining a partition and recomputing centroids within each refined subgroup cannot increase the within-group variance. Therefore, revealing additional prefix tokens progressively reduces semantic ambiguity in expectation.

For numerical stability in multi-target training, we normalize the weights so that their sum equals the ID length $L$:
\[
w_k^{\mathrm{fg}} \leftarrow \frac{w_k^{\mathrm{fg}}}{\sum_{j=1}^{L} w_j^{\mathrm{fg}}}\cdot L.
\]
This length-preserving normalization keeps the scale of the Front-Greater objective comparable to standard cross-entropy, preventing gradient imbalance and enabling stable joint learning with other token-weighting objectives.

This strategy aligns naturally with generative recommendation using semantic IDs, where early tokens encode coarse semantics under RQ-VAE’s hierarchical structure and are often used in top-$k$ decoding or candidate pruning~\cite{tiger}, making their accurate prediction particularly critical.

\subsubsection{\textbf{Frequency Token Weighting}}

As shown in Figure~\ref{fig:overrecommend}, generative models such as T5~\cite{t5} are prone to popularity bias, disproportionately favoring frequently occurring items and tokens. 
From an information-theoretic perspective, rare tokens carry higher discriminative value, and many hard negatives differ from the target item in only one or two rare tokens. Effectively distinguishing such cases requires the model to devote more learning capacity to rare tokens rather than allowing frequent tokens to dominate training.

Motivated by these observations, we propose \textbf{Frequency Token Weighting}, which assigns larger weights to tokens that appear less frequently in the dataset. Our formulation draws inspiration from the concept of \textbf{Effective Number of Samples} proposed by \citet{effectiveNum}, which defines the amount of information contributed by repeated observations of a class. For a token that appears $n$ times, the effective number is defined as:
\[
E_n = \frac{1 - \beta^n}{1 - \beta},
\]
where $\beta \in (0,1)$ controls the sensitivity to frequency. This formulation corresponds to a truncated geometric series, suggesting diminishing marginal information as $n$ increases. We formalize this property in the following lemma.

\begin{lemma}[Token Frequency Induces Diminishing Marginal Information Gain]
\label{lemma:freq}
Let 
$T$ be a token that appears $n$ times in the training data, and assume the cumulative information contributed by repeated observations of $T$ follows the effective number $E_n$. Then $E_n$ is non-decreasing and concave in $n$, and the marginal information gain contributed by the $n$-th occurrence,
\[
\Delta_n = E_n - E_{n-1} = \beta^{n-1},
\]
is strictly decreasing with respect to 
$n$. Consequently, each additional occurrence of a frequent token contributes less marginal information than earlier occurrences.
\end{lemma}
Lemma~\ref{lemma:freq} shows that the marginal information gain of the $n$-th occurrence decreases monotonically as $n$ increases. This implies that additional occurrences of frequent tokens provide increasingly redundant training signal, whereas rare tokens contribute stronger marginal learning signal. 
Based on this result, we define the frequency-based token weight as
\[
w^{\mathrm{fr}}_k \propto \max(\frac{1}{E_{n_k}},0),
\]
where $n_k$ denotes the frequency of the $k$-th token in the dataset. This weighting scheme emphasizes token occurrences with higher expected marginal information gain, while downweighting frequent tokens whose contributions are largely redundant. Similar to the Front-Greater strategy, we normalize the weights such that the sum of token weights equals the semantic ID sequence length, ensuring training stability and preserving the overall loss scale.

\subsection{Multi-Target Framework with Curriculum Learning}
\label{sec:curr-aware weighting scheme}
Our generative recommender jointly optimizes three complementary token-level objectives: (1) Front-Greater Token Loss ($L_{\text{fg}}$), (2) Frequency Token Loss ($L_{\text{fr}}$), and (3) the standard cross-entropy loss ($L_{\text{or}}$). Each objective operates on the negative log-likelihood of predicting the next token. For the $i$-th token, we expand the softmax and the weighted loss takes the form
\[
\ell_{i}^{\text{weighted}}
= 
- w_i
\log
\left(
\frac{\exp(f_\theta^{y_i}(x, y_{<i}))}
     {\sum_{v \in \textit{V}} \exp(f_\theta^{v}(x, y_{<i}))}
\right),
\]
where $w_i$ is the token weight, 
$f_\theta^{v}(x, y_{<i})$ is the logit assigned to token $v$ at step $i$, and $\textit{V}$ denotes the model's token vocabulary.

\subsubsection{\textbf{Incorporating Multi-Target Scaling Parameters}}
To combine the three objectives in a principled manner, we adopt objective-dependent scaling parameters \cite{kendall2018multi}. 
For each objective $j \in \{\text{fg},\,\text{fr},\,\text{or}\}$, we introduce a learnable parameter $\lambda_j$ that
parameterizes the relative contribution of each objective.
These parameters are normalized to obtain objective weights
\[
\alpha_j
=
\frac{\lambda_j}{\lambda_{\text{fg}} + \lambda_{\text{fr}} + \lambda_{\text{or}}},
\]
which lie on the probability simplex and are used to modulate the learning dynamics.
Specifically, the normalized weight $\alpha_j$ is applied as a scaling factor on the logits before the softmax operation, effectively controlling the sharpness of the token probability distribution associated with each objective.
Taking Front-Greater Token Weighting as an example, scaling the logits by $\alpha_{\text{fg}}$ and substituting into the likelihood gives:
\[
\ell^{\text{fg-scale}}_i
=
- w_i^{\text{fg}}
\log
\left(
\frac{\exp(\alpha_{\text{fg}} f_\theta^{y_i})}
     {\sum_{v} \exp(\alpha_{\text{fg}} f_\theta^{v})}
\right).
\]
Rewriting the expression, we obtain:
\begin{align*}
\ell^{\text{fg-scale}}_i
&=
\alpha_{\text{fg}} \ell^{\text{fg}}_i
+
w_i^{\text{fg}}
\log
\left(
\frac{
\sum_v \exp(\alpha_{\text{fg}} f_\theta^{v})
}{
\left( \sum_v \exp(f_\theta^{v}) \right)^{\alpha_{\text{fg}}}
}
\right).
\end{align*}
The second logarithmic term is independent of the correct class token, but depends on $\alpha_{\text{fg}}$ and the model logits through the normalization. Under the limiting cases $\alpha_{\text{fg}} \to 1$ and $\alpha_{\text{fg}} \to 0$ (a behavior encouraged by the curriculum schedule introduced in Section~\ref{sec:curriculum}), this term collapses to a constant (e.g., $w_i^{\text{fg}}\log \textit{V}$ or $0$) and can be approximately ignored.
Thus, the scaled objective simplifies to:
\[
\ell^{\text{fg-scale}}_i = \alpha_{\text{fg}} \ell^{\text{fg}}_i.
\]
Summing over all positions yields the overall Front-Greater Token Loss:
\[
L_{\text{fg-scale}} = \alpha_{\text{fg}} L_{\text{fg}}.
\]
By the same derivation, the other objectives become:
\[
L_{\text{fr-scale}} = \alpha_{\text{fr}} L_{\text{fr}},\ \text{and}
\qquad
L_{\text{or-scale}} = \alpha_{\text{or}} L_{\text{or}}.
\]
\begin{table}[t]
\centering
\small
\setlength{\tabcolsep}{5.5pt}
\begin{tabular}{lcccc}
\toprule
Statistic & Music. & Industr. & Yelp & MovieLens1M \\
\midrule
Total Interactions & 511836 & 412947 & 316354 & 1000209 \\
User Number        & 57439 & 50985 & 30431  & 6040    \\
Number of Items    & 24587 & 25847 & 20033  & 3883    \\
History Mean Length & 8.91 & 8.10  & 10.40  & 165.60  \\
\bottomrule
\end{tabular}
\caption{Statistics of the benchmark datasets.}
\label{tab:dataset-statistics}
\end{table}
\newcommand{\sigstar}{\llap{\textsuperscript{*}}}
\newcommand{\mcell}[1]{\multicolumn{1}{S[table-format=1.4]}{#1}}

\newcommand{\sbest}[1]{%
  \multicolumn{1}{S[table-format=1.4,parse-numbers=false,mode=text]}{%
    \underline{\textnormal{#1}}%
  }%
}

\newcommand{\best}[1]{%
  \multicolumn{1}{S[table-format=1.4,parse-numbers=false,mode=text]}{%
    \textbf{#1}\textsuperscript{*}%
  }%
}

\newcommand{\hdr}[1]{\multicolumn{1}{c}{#1}}

\begin{table*}[t]
\centering
\small
\setlength{\tabcolsep}{2.1pt}
\renewcommand{\arraystretch}{1.05}

\begin{tabular}{ll *{16}{S[table-format=1.4]}}
\toprule
& & \multicolumn{4}{c}{\textbf{Musical Instruments}}
& \multicolumn{4}{c}{\textbf{Industrial and Scientific}}
& \multicolumn{4}{c}{\textbf{Yelp}}
& \multicolumn{4}{c}{\textbf{MovieLens 1M}} \\
\cmidrule(lr){3-6} \cmidrule(lr){7-10} \cmidrule(lr){11-14} \cmidrule(lr){15-18}

\textbf{Category} & \textbf{Method}
& \hdr{H@5} & \hdr{N@5} & \hdr{H@10} & \hdr{N@10}
& \hdr{H@5} & \hdr{N@5} & \hdr{H@10} & \hdr{N@10}
& \hdr{H@5} & \hdr{N@5} & \hdr{H@10} & \hdr{N@10}
& \hdr{H@5} & \hdr{N@5} & \hdr{H@10} & \hdr{N@10} \\
\midrule

\multirow{2}{*}{Traditional}
& GRU4Rec
& 0.0193 & 0.0129 & 0.0292 & 0.0161
& 0.0162 & 0.0110 & 0.0231 & 0.0132
& 0.0143 & 0.0093 & 0.0232 & 0.0121
& 0.0993 & 0.0574 & 0.1806 & 0.0835 \\

& SASRec
& 0.0250 & 0.0158 & 0.0400 & 0.0207
& 0.0179 & 0.0116 & 0.0295 & 0.0153
& 0.0166 & 0.0099 & 0.0279 & 0.0137
& 0.1108 & 0.0648 & 0.1902 & 0.0904 \\

\midrule
\multirow{7}{*}{\makecell[l]{Token-weighted\\ GR}}
& TIGER
& 0.0316 & 0.0203 & 0.0501 & 0.0263
& 0.0246 & 0.0158 & 0.0393 & 0.0206
& 0.0240 & 0.0155 & 0.0391 & 0.0202
& 0.1859 & 0.1275 & 0.2648 & 0.1530 \\

& TIGER + Rank
& 0.0311 & 0.0204 & 0.0482 & 0.0258
& 0.0233 & 0.0151 & 0.0381 & 0.0199
& 0.0241 & 0.0155 & 0.0396 & 0.0205
& 0.1841 & 0.1268 & 0.2644 & 0.1528 \\

& TIGER + Pos
& 0.0320 & 0.0207 & 0.0497 & 0.0264
& 0.0248 & 0.0161 & 0.0396 & \sbest{0.0208}
& 0.0239 & 0.0154 & 0.0387 & 0.0201
& 0.1868 & \sbest{0.1302} & \sbest{0.2687} & \sbest{0.1566} \\

& TIGER + CFT
& 0.0321 & 0.0208 & 0.0502 & 0.0266
& 0.0244 & 0.0157 & 0.0387 & 0.0203
& 0.0239 & 0.0154 & 0.0390 & 0.0202
& 0.1876 & 0.1299 & 0.2665 & 0.1553 \\

& TIGER + IGD
& \sbest{0.0325} & \sbest{0.0210} & \sbest{0.0503} & \sbest{0.0268}
& \sbest{0.0251} & \sbest{0.0161} & \sbest{0.0396} & 0.0207
& \sbest{0.0245} & \sbest{0.0158} & \sbest{0.0401} & \sbest{0.0208}
& \sbest{0.1878} & 0.1296 & 0.2617 & 0.1534 \\

& TIGER + Ours
& \best{0.0330} & \best{0.0215} & \best{0.0512} & \best{0.0273}
& \best{0.0255} & \best{0.0164} & \best{0.0398} & \best{0.0210}
& \best{0.0264} & \best{0.0175} & \best{0.0419} & \best{0.0225}
& \best{0.1945} & \best{0.1354} & \best{0.2727} & \best{0.1605} \\

\midrule
\multicolumn{2}{l}{\textit{Improve. (vs. TIGER)}} 
& \multicolumn{1}{c}{\textit{6.1\%}} & \multicolumn{1}{c}{\textit{5.6\%}} & \multicolumn{1}{c}{\textit{2.2\%}} & \multicolumn{1}{c}{\textit{3.8\%}}
& \multicolumn{1}{c}{\textit{3.72\%}} & \multicolumn{1}{c}{\textit{3.48\%}} & \multicolumn{1}{c}{\textit{1.26\%}} & \multicolumn{1}{c}{\textit{2.01\%}}
& \multicolumn{1}{c}{\textit{10.0\%}} & \multicolumn{1}{c}{\textit{13.3\%}} & \multicolumn{1}{c}{\textit{7.2\%}} & \multicolumn{1}{c}{\textit{11.4\%}}
& \multicolumn{1}{c}{\textit{4.6\%}} & \multicolumn{1}{c}{\textit{6.2\%}} & \multicolumn{1}{c}{\textit{3.0\%}} & \multicolumn{1}{c}{\textit{4.9\%}} \\

\bottomrule
\end{tabular}
\caption{Main results on four benchmark datasets. Traditional recommenders are compared with token-weighted generative recommenders (GR) based on TIGER. Best results are in \textbf{bold}, second-best are \underline{underlined}. * indicates a statistically significant difference ($p < 0.05$) from the best baseline (i.e. best compared methods).}
\label{tab:main_exp}
\end{table*}

\subsubsection{\textbf{Curriculum Learning Over Objective Weights}}
\label{sec:curriculum}
While Front-Greater and Frequency weighting provide complementary training signals, their effects are most useful at different stages of optimization.
Early in training, the model benefits from emphasizing (1) Front-Greater Token Loss, which stabilizes early-token prediction, and (2) the original cross-entropy loss, which promotes general learning robustness.
In contrast, once the model has learned a reasonable prefix structure, shifting emphasis to Frequency Token Loss enables it to focus on rare, discriminative tokens, which in turn improves fine-grained item separation and alleviates popularity bias.

To encode this dynamic behavior, we introduce a curriculum schedule over objective parameters using an exponential decay function:
\[
e^{-ct},
\]
where $t$ is the training step and $c>0$ controls how quickly the curriculum transitions.
We define curriculum-adjusted objective parameters:
\[
\lambda_{\text{fg}}' = e^{-ct} \lambda_{\text{fg}},
\qquad
\lambda_{\text{or}}' = e^{-ct} \lambda_{\text{or}},\ \text{and}
\qquad
\lambda_{\text{fr}}' = (1 - e^{-ct}) \lambda_{\text{fr}}.
\]
Under this formulation, Front-Greater and Original CE losses dominate early training, while Frequency Token Loss gradually takes over as $t$ increases. We then compute the mixing coefficients:
\[
\alpha_j'
=
\frac{\lambda_j'}
     {\lambda_{\text{fg}}' + \lambda_{\text{fr}}' + \lambda_{\text{or}}'}.
\]

Finally, the multi-target training objective with curriculum learning is:
\[
L
=
\alpha_{\text{fg}}' L_{\text{fg}}
+
\alpha_{\text{fr}}' L_{\text{fr}}
+
\alpha_{\text{or}}' L_{\text{or}}.
\]
This formulation enables the model to first focus on learning stable front-position semantics and general predictive ability, and later adapt its training emphasis toward rare-token discrimination. The resulting curriculum thus aligns naturally with the learning dynamics desired in generative recommendation.
\section{Experiment}
We describe the experimental setup, including datasets, evaluation protocol, compared methods, and implementation details, followed by empirical analysis of our research questions.

\subsection{Experimental Setup}
\subsubsection{\textbf{Datasets and Evaluation Settings}}
We evaluate on four widely used datasets covering different recommendation scenarios: \textit{Musical Instruments} and \textit{Industrial and Scientific} from Amazon~\cite{amazon23}, \textit{Yelp}\footnote{Yelp dataset available at \url{https://business.yelp.com/data/resources/open-dataset/}}, and \textit{MovieLens 1M}~\cite{movielens}.  
\textit{Musical Instruments} and \textit{Industrial and Scientific} represent large-scale e-commerce data with many cold-start items, \textit{Yelp} reflects real-world user--business interactions with moderate sparsity, and \textit{MovieLens 1M} is a dense benchmark with long user histories. Table~\ref{tab:dataset-statistics} summarizes the statistics of all datasets used in our experiments.

We apply standard five-core\footnote{Items and users with fewer than five interactions are removed from the dataset.} filtering and adopt a leave-one-out\footnote{The last interaction is used for testing, the second-to-last for validation, and the others for training.} protocol for sequential recommendation. Performance is evaluated using Hit@$K$ and NDCG@$K$, which measure top-$K$ accuracy and ranking quality, respectively.

\subsubsection{\textbf{Compared Methods}}
We compare our method with representative sequential recommenders and token-weighting methods. For fairness, all token-level methods are implemented on the TIGER architecture~\cite{tiger}, which combines RQ-VAE semantic IDs with a T5 backbone.

\begin{itemize}
    \item \textbf{GRU4Rec}~\cite{gru4rec}: A recurrent model that captures short-term sequential patterns using GRUs.
    \item \textbf{SASRec}~\cite{sasrec}: A transformer-based sequential recommender modeling long-range dependencies via self-attention.
    \item \textbf{Rank}~\cite{letter}: A ranking-guided loss from LETTER that sharpens the softmax distribution to emphasize hard negatives.
    \item \textbf{Pos}~\cite{cft}: A positional weighting baseline that assigns larger weights to earlier tokens in the generated sequence.
    \item \textbf{CFT}~\cite{cft}: A causal fine-tuning method that reweights tokens based on their estimated causal contribution to item prediction.
    \item \textbf{IGD}~\cite{igd}: An information-theoretic approach that emphasizes tokens with high information gain about the target item.
\end{itemize}

\subsubsection{\textbf{Implementation Details}}
\label{sec:implement}
All experiments are conducted on two NVIDIA RTX 3090 GPUs and one RTX A6000 GPU, with each method run five times using the same data splits. We report average performance and assess statistical significance ($p<0.05$). 

For TIGER-based models, semantic item IDs are constructed using an RQ-VAE with four codebooks of size 256 each.\footnote{ To prevent identifier collisions, any duplicate IDs are resolved by randomly reassigning the final token, ensuring that each item is associated with a unique identifier.} Item embeddings are extracted using the same semantic encoder (LLaMA2-7B~\cite{llama}) across all methods. $\beta$ in frequency weighting is set to 0.99.
All other hyperparameters for baselines follow their official implementations.

\begin{table}[t]
\centering
\footnotesize
\setlength{\tabcolsep}{5pt}
\renewcommand{\arraystretch}{1.05}

\begin{tabular}{ll *{4}{S[table-format=1.4]}}
\toprule
Dataset & Model
& \hdr{HIT@5} & \hdr{N@5} & \hdr{HIT@10} & \hdr{N@10} \\
\midrule

\multirow{5}{*}{Music Instruments}
& TIGER
& 0.0316 & 0.0203 & 0.0501 & 0.0263 \\
& + FrontGreater
& 0.0323 & 0.0209 & 0.0501 & 0.0265 \\
& + Frequency
& 0.0324 & 0.0211 & 0.0500 & 0.0268 \\
& + Multi-Target
& \sbest{0.0327} & \sbest{0.0212} & \sbest{0.0509} & \sbest{0.0271} \\
& + Curriculum
& \best{0.0330} & \best{0.0215} & \best{0.0512} & \best{0.0273} \\

\midrule
\multirow{5}{*}{Yelp}
& TIGER
& 0.0240 & 0.0155 & 0.0391 & 0.0202 \\
& + FrontGreater
& 0.0251 & \sbest{0.0166} & 0.0401 & 0.0214 \\
& + Frequency
& 0.0253 & 0.0162 & 0.0401 & 0.0210 \\
& + Multi-Target
& \sbest{0.0255} & 0.0165 & \sbest{0.0415} & \sbest{0.0216} \\
& + Curriculum
& \best{0.0264} & \best{0.0175} & \best{0.0419} & \best{0.0225} \\

\midrule
\multirow{5}{*}{MovieLens1M}
& TIGER
& 0.1854 & 0.1293 & 0.2658 & 0.1551 \\
& + FrontGreater
& 0.1891 & 0.1303 & \sbest{0.2713} & 0.1568 \\
& + Frequency
& 0.1914 & 0.1319 & 0.2697 & 0.1572 \\
& + Multi-Target
& \sbest{0.1928} & \sbest{0.1336} & 0.2680 & \sbest{0.1577} \\
& + Curriculum
& \best{0.1945} & \best{0.1354} & \best{0.2727} & \best{0.1605} \\

\bottomrule
\end{tabular}

\caption{Ablation study on token-weighting strategies and curriculum-guided multi-target learning. Components are added progressively to demonstrate their incremental contributions. * indicates a statistically significant improvement over the second-best result ($p < 0.05$).}
\label{tab:ablation}
\end{table}
\subsection{Experiment Results}

We answer the following research questions through controlled experiments:

\begin{itemize}
    \item \textbf{RQ1}: Does our method improve the performance of generative recommenders compared with strong token-weighting baselines?
    \item \textbf{RQ2}: How does each component of our method contribute to the overall performance gain?
    \item \textbf{RQ3}: Is our method effective across different semantic-ID constructions and model scales?
    \item \textbf{RQ4}: Does our method provide more robust recommendations in practice?
    \item \textbf{RQ5}: How does our method perform across head and tail items under frequency-based data splits?
    \item \textbf{RQ6}: How does our method perform under different speeds of curriculum progression?
\end{itemize}
\subsubsection{\textbf{Main Results (RQ1)}}

We first evaluate whether our proposed method yields performance improvements across the four datasets with varying domain characteristics. The detailed results are reported in Table~\ref{tab:main_exp}. The key observations are summarized as the following points. (1) Our method consistently improves performance across all datasets.  On average, we observe gains of 6.11\% (Hit@5), 7.14\% (NDCG@5), 3.39\% (Hit@10), and 5.52\% (NDCG@10) over the base TIGER model. The improvement is most pronounced on the \textit{Yelp} dataset, where our method achieves a 10.0\% increase in Hit@5 and a 13.3\% increase in NDCG@5. These results demonstrate the robustness and generality of our method across domains of differing sparsity and sequence lengths. (2) Our method outperforms all token-weighting baselines, including Pos, CFT, IGD, and the traditional recommenders. Although IGD achieves the closest performance to ours, which reflects the shared intuition that tokens with higher information gain should be emphasized, it measures information gain through entropy under a single objective. In contrast, our method is specifically designed for generative recommendation with semantic IDs, which exploits the hierarchical structure and long-tailed distribution of semantic IDs within a  multi-target framework with curriculum learning. This richer learning framework likely contributes to the superior and more stable performance of our method across all datasets and metrics.

\begin{table}[!t]
\centering
\footnotesize
\setlength{\tabcolsep}{5pt}
\renewcommand{\arraystretch}{1.05}

\begin{tabular}{ll *{4}{S[table-format=1.4]}}
\toprule
Dataset & Model
& \hdr{HIT@5} & \hdr{N@5} & \hdr{HIT@10} & \hdr{N@10} \\
\midrule

\multirow{3}{*}{Music Instruments}
& TIGER
& 0.0304 & 0.0198 & 0.0466 & 0.0251 \\
& + ours
& \best{0.0315} & \best{0.0206} & \best{0.0485} & \best{0.0261} \\
& \textit{Improvement}
& \multicolumn{1}{c}{\textit{3.61\%}}
& \multicolumn{1}{c}{\textit{4.04\%}}
& \multicolumn{1}{c}{\textit{4.07\%}}
& \multicolumn{1}{c}{\textit{3.98\%}} \\

\midrule
\multirow{3}{*}{Yelp}
& TIGER
& 0.0190 & 0.0121 & 0.0310 & 0.0159 \\
& + ours
& \best{0.0212} & \best{0.0140} & \best{0.0346} & \best{0.0183} \\
& \textit{Improvement}
& \multicolumn{1}{c}{\textit{11.57\%}}
& \multicolumn{1}{c}{\textit{15.7\%}}
& \multicolumn{1}{c}{\textit{11.61\%}}
& \multicolumn{1}{c}{\textit{17.3\%}} \\

\midrule
\multirow{3}{*}{MovieLens1M}
& TIGER
& 0.1702 & 0.1170 & 0.2477 & 0.1421 \\
& + ours
& \best{0.1790} & \best{0.1222} & \best{0.2544} & \best{0.1464} \\
& \textit{Improvement}
& \multicolumn{1}{c}{\textit{5.17\%}}
& \multicolumn{1}{c}{\textit{4.44\%}}
& \multicolumn{1}{c}{\textit{2.70\%}}
& \multicolumn{1}{c}{\textit{3.02\%}} \\

\bottomrule
\end{tabular}

\caption{Performance comparison of TIGER+Ours vs. original TIGER with semantic IDs from Product-Quantization (PQ) on Musical Instruments (Amazon'23), Yelp, and MovieLens 1M. * indicates a statistically significant improvement over the result of original TIGER ($p < 0.05$).}
\label{tab:pq}
\end{table}




\begin{table}[!t]
\centering
\footnotesize
\setlength{\tabcolsep}{5pt}
\renewcommand{\arraystretch}{1.05}

\begin{tabular}{ll *{4}{S[table-format=1.4]}}
\toprule
Dataset & Model
& \hdr{HIT@5} & \hdr{N@5} & \hdr{HIT@10} & \hdr{N@10} \\
\midrule

\multirow{3}{*}{Music Instruments}
& TIGER
& 0.0290 & 0.0195 & 0.0408 & 0.0233 \\
& + ours
& \best{0.0302} & \best{0.0204} & \best{0.0425} & \best{0.0243} \\
& \textit{Improvement}
& \multicolumn{1}{c}{\textit{4.30\%}}
& \multicolumn{1}{c}{\textit{4.09\%}}
& \multicolumn{1}{c}{\textit{4.42\%}}
& \multicolumn{1}{c}{\textit{4.26\%}} \\

\midrule
\multirow{3}{*}{Industrial and Scientific}
& TIGER
& 0.0244 & 0.0162 & 0.0352 & 0.0197 \\
& + ours
& \best{0.0253} & \best{0.0168} & \best{0.0365} & \best{0.0205} \\
& \textit{Improvement}
& \multicolumn{1}{c}{\textit{5.17\%}}
& \multicolumn{1}{c}{\textit{4.44\%}}
& \multicolumn{1}{c}{\textit{2.70\%}}
& \multicolumn{1}{c}{\textit{3.02\%}} \\

\midrule
\multirow{3}{*}{Yelp}
& TIGER
& 0.0375 & 0.0277 & 0.0433 & 0.0296 \\
& + ours
& \best{0.0388} & \best{0.0284} & \best{0.0457} & \best{0.0307} \\
& \textit{Improvement}
& \multicolumn{1}{c}{\textit{3.47\%}}
& \multicolumn{1}{c}{\textit{5.54\%}}
& \multicolumn{1}{c}{\textit{2.52\%}}
& \multicolumn{1}{c}{\textit{3.72\%}} \\


\bottomrule
\end{tabular}

\caption{Performance comparison of TIGER+Ours vs. original TIGER on LLaMA3.2-3b  on Musical Instruments (Amazon'23), Industrial and Scientific (Amazon'23), and Yelp. * indicates a statistically significant improvement over the result of original TIGER ($p < 0.05$).}

\label{tab:llm}
\end{table}



\begin{table}[!t]
\centering
\footnotesize
\setlength{\tabcolsep}{5pt}
\renewcommand{\arraystretch}{1.05}

\begin{tabular}{ll *{4}{S[table-format=1.4]}}
\toprule
Dataset & Model
& \hdr{Top-4} & \hdr{Top-3} & \hdr{Top-2} & \hdr{Top-1} \\
\midrule

\multirow{3}{*}{Music Instruments}
& TIGER
& 0.0206 & 0.0266 & 0.0427 & 0.1162 \\
& + FrontGreater
& 0.0210 & 0.0270 & 0.0438 & 0.1183 \\
& + Ours
& \best{0.0215} & \best{0.0277} & \best{0.0448} & \best{0.1193} \\

\midrule
\multirow{3}{*}{Yelp}
& TIGER
& 0.0155 & 0.0157 & 0.0178 & 0.0665 \\
& + FrontGreater
& 0.0166 & 0.0170 & 0.0189 & 0.0675 \\
& + Ours
& \best{0.0175} & \best{0.0179} & \best{0.0201} & \best{0.0697} \\

\midrule
\multirow{3}{*}{MovieLens1M}
& TIGER
& 0.1275 & 0.1283 & 0.1318 & 0.2044 \\
& + FrontGreater
& 0.1304 & 0.1311 & 0.1351 & 0.2097 \\
& + Ours
& \best{0.1354} & \best{0.1362} & \best{0.1395} & \best{0.2104} \\

\bottomrule
\end{tabular}

\caption{Performance comparison (NDCG@5) of the original TIGER and TIGER+Ours with top-$k$ tokens aligned to those of the target item on Musical Instruments (Amazon'23), Yelp, and MovieLens~1M.
Top-4 results correspond to the original evaluation setting.
* indicates a statistically significant improvement over the original TIGER ($p < 0.05$).}
\label{tab:topk}
\end{table}

\subsubsection{\textbf{Ablation Study (RQ2)}}
To understand the contribution of each component in our framework, we conduct an ablation study in which we progressively introduce (1) Front-Greater Token Weighting, (2) Frequency Token Weighting, (3) adaptive multi-target learning, and (4) curriculum learning. For each configuration, we train the generative model independently and evaluate its performance across all datasets. The results are summarized in Table~\ref{tab:ablation}.

We first examine the \textbf{Front-Greater} setting, where we apply only the Front-Greater Token Weighting and use a simple heuristic linear schedule to gradually decrease its influence over training, eventually reverting to standard cross-entropy. Even with this minimal setup, the model already shows noticeable improvements, demonstrating that emphasizing early-position tokens provides meaningful learning signals.

Next, in the \textbf{Front-Greater + Frequency} setting, we incorporate the Frequency Token Weighting strategy. Similar to the first configuration, we use a linear schedule that transitions from Front-Greater to standard cross-entropy and finally to Frequency-based weighting. This setup consistently improves performance over the baseline, but the results are not entirely stable.


To address this limitation, we introduce a \textbf{multi-target learning} framework that jointly optimizes all objectives and adaptively learns their mixing ratios via learnable scaling parameters $\lambda_j$, (normalized to $\alpha_j$), yielding an overall objective $\sum \alpha_j L_j$, rather than relying on predefined schedules. Compared to heuristic linear schedules, this approach produces more stable and consistently stronger results, highlighting the benefit of dynamically adjusting the relative importance of each objective during training.

Finally, we incorporate a \textbf{curriculum} into the multi-target framework, which introduces an explicit prior over training dynamics. Specifically, the curriculum encourages the model to focus on the more stable objectives, which are Front-Greater and standard cross-entropy, during early training, and gradually shift attention toward the more fine-grained Frequency objective.
With the curriculum-guided multi-target training scheme, the model achieves the best performance across all datasets, highlighting the effectiveness and robustness of combining adaptive weighting with curriculum guidance.

\subsubsection{\textbf{Generalizability (RQ3)}}

We first investigate whether our method generalizes beyond semantic IDs generated by Residual Quantization (RQ). Specifically, we evaluate our framework using Product Quantization (PQ)–based IDs, a widely adopted alternative for constructing compact semantic representations \cite{pq, pqcf, vqpq}. For a fair comparison, we keep all model components identical to the RQ-based setting and replace only the ID generation module. To match the four-level codebooks used in RQ-VAE, we divide each item’s semantic embedding into four equal segments. Each segment is quantized with an independently learned PQ codebook.
As shown in Table~\ref{tab:pq}, our method consistently improves performance across all datasets even under PQ-based IDs. On average, we observe gains of 6.78\% in Hit@5, 8.06\% in NDCG@5, 6.13\% in Hit@10, and 8.1\% in NDCG@10, which are comparable to those obtained under the RQ setting. 
Since PQ-based IDs form an ordered token sequence whose prefixes induce progressively refined partitions of the item space, Front-Greater weighting remains effective by emphasizing tokens that yield larger prefix-conditional dispersion reduction. The remaining components, including Frequency weighting and the multi-target curriculum, are likewise agnostic to the ID construction mechanism.

We further evaluate the generalizability of our framework across different backbone model scales. While the main experiments in Table~\ref{tab:main_exp} are conducted on TIGER with T5~\cite{t5}, we additionally replace the backbone with LLaMA3.2-3B~\cite{llama3}. 
To enable efficient finetuning, we employ 8-bit LoRA~\cite{lora} for both training and inference, with the rank parameter set to 8, the scaling factor $\alpha$ set to 32, and the dropout rate set to 0.05.
As shown in Table~\ref{tab:llm}, our method consistently improves performance across datasets under the larger backbone model.

These results demonstrate that our framework is not tied to a specific quantization method or language model architecture. Instead, it generalizes well across different semantic-ID representations, model scales, and recommendation domains.
\definecolor{ForestGreen}{RGB}{34,139,34}
\definecolor{BrickRed}{RGB}{178,34,34}
\definecolor{SunsetOrange}{RGB}{255,99,71}

\begin{table}[t]
\centering
\scriptsize
\setlength{\tabcolsep}{2pt}
\renewcommand{\arraystretch}{1.05}
\begin{tabular}{p{2cm} p{3cm} p{3cm}}
\toprule
\textbf{Ground Truth} & \textbf{Ours (Top-3)} & \textbf{TIGER (Top-3)} \\
\midrule
\textcolor{BrickRed}{Snark SN-8 Super Tight All Instrument Tuner}
&
\begin{enumerate}[leftmargin=*, nosep, topsep=0pt]
  \item \textcolor{ForestGreen}{D'Addario Accessories Guitar Tuner - Micro Headstock Tuner (Clip-on)}
  \item D'Addario Guitar Strings - Phosphor Bronze Acoustic (EJ41)
  \item \textcolor{BrickRed}{Snark SN-8 Super Tight All Instrument Tuner}
\end{enumerate}
&
\begin{enumerate}[leftmargin=*, nosep, topsep=0pt]
  \item \textcolor{SunsetOrange}{Inlay Sticker Fret Markers (Side Marker Dots)}
  \item \textcolor{SunsetOrange}{Inlaystickers Sticker/Decal (F-020PC-WT)}
  \item \textcolor{SunsetOrange}{Inlaystickers Sticker/Decal (B-143HB-14)}
\end{enumerate}
\\
\bottomrule
\end{tabular}
\vspace{2mm}
\caption{Qualitative example on \textbf{Musical Instruments (Amazon'23)}. We show the ground-truth next item, top-3 predictions from our method and the original TIGER model. \textcolor{BrickRed}{Red} indicates the target item, \textcolor{ForestGreen}{green} denotes items semantically related to the target, and \textcolor{SunsetOrange}{orange} marks items unrelated to the target.}
\label{tab:qual_example_music}
\vspace{-2mm}
\end{table}

\begin{figure*}[t]
    \centering
    \includegraphics[width=0.95\linewidth]{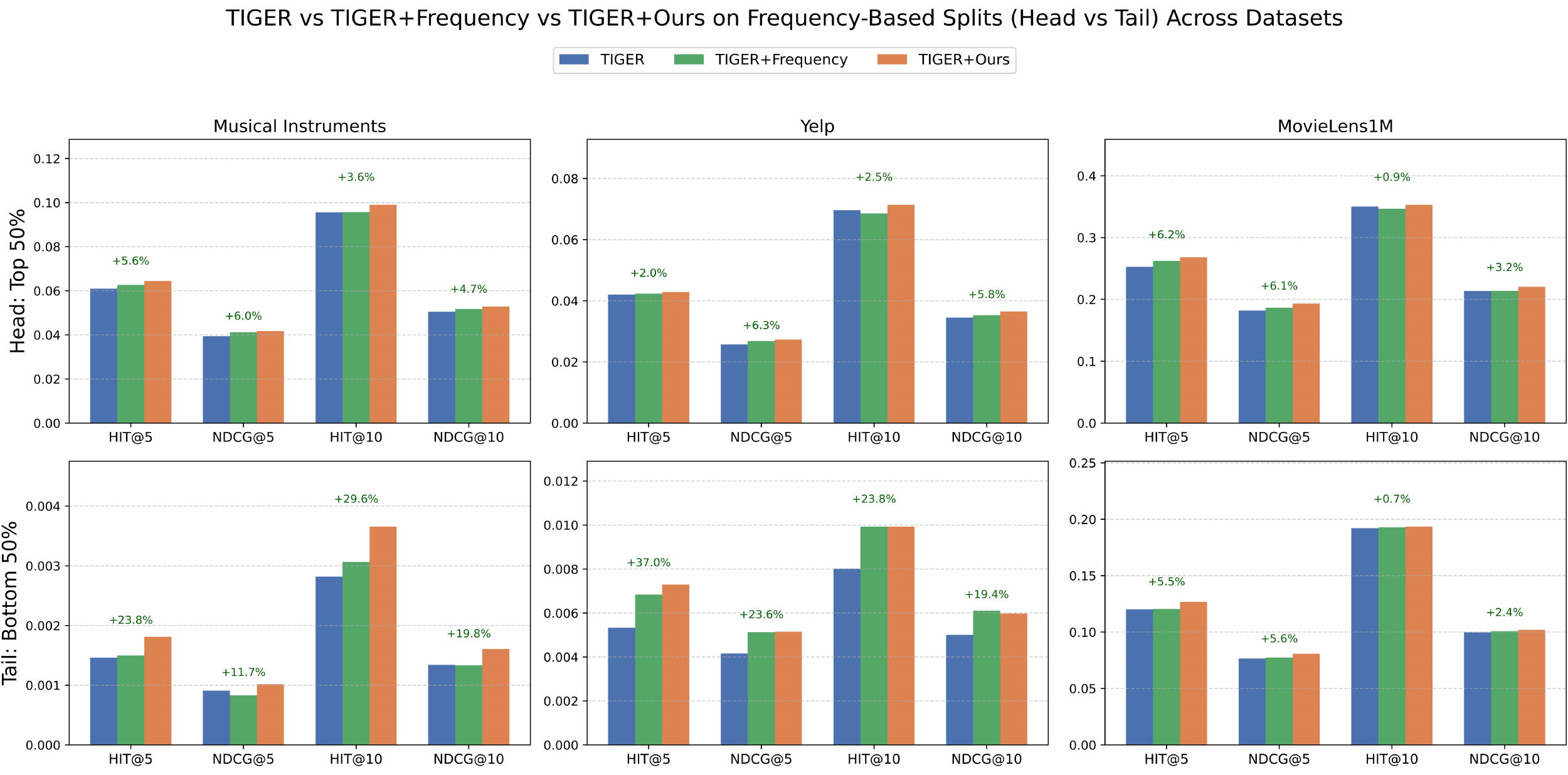}
    \caption{Head--tail evaluation based on item frequency in the test set. Items are split into head (top 50\%) and tail (bottom 50\%) groups by interaction frequency. Our method consistently outperforms TIGER on both sets across all datasets, with particularly strong improvements on tail items, highlighting the effectiveness of frequency-based token weighting in addressing long-tailed recommendation challenges.}
    \label{fig:headtail}
\end{figure*}

\subsubsection{\textbf{Robustness (RQ4)}}
We further evaluate the robustness of our method under a more realistic evaluation setting. In practice, requiring a model to exactly predict the next item is often overly strict, especially for generative recommenders based on RQ-based semantic IDs, where tokens encode semantics from coarse-grained to fine-grained levels. Motivated by this property, we consider predictions that match the target item on the top-$k$ tokens as partially correct, reflecting semantic similarity even when the full ID is not perfectly matched. The quantitative results are shown in Table~\ref{tab:topk}. Across all evaluation schemes, our method consistently outperforms the original TIGER model, demonstrating stronger robustness under relaxed semantic matching criteria.

In addition, we provide a qualitative example to illustrate the practical advantages of our method. As shown in Table~\ref{tab:qual_example_music}, our top-3 predictions include the ground-truth target item, whereas the target item does not appear in the predictions of the original TIGER. Although our model does not rank the target item at the first position, it places a semantically related item (i.e., another tuner) at the top of the list. This behavior is consistent with the user’s interaction history, which mainly consists of audio and instrument-related equipment (e.g., stands, microphones, and audio interfaces), suggesting that tuner-related accessories are plausible next items. In contrast, the original TIGER ranks three unrelated items (stickers) at the top, indicating a severe semantic mismatch. This example highlights that our method produces recommendations that are not only quantitatively superior but also qualitatively more robust and better aligned with user intent, which is an aspect not fully captured by standard exact-match evaluation metrics in sequential recommendation.

\subsubsection{\textbf{Head--Tail Analysis (RQ5)}}
To further assess robustness under frequency imbalance, we conduct a head–tail analysis based on item popularity. Items in the test set are ranked by interaction frequency and split into two equal-sized groups: a \textit{head} set (top 50\% most frequent items) and a \textit{tail} set (bottom 50\% least frequent items). Model performance is evaluated separately on these two subsets. Figure~\ref{fig:headtail} compares the original TIGER model, TIGER with Frequency weighting only, and TIGER with our full method. Across all datasets, our method consistently outperforms TIGER on both head and tail sets, with substantially larger gains observed on the tail set. This highlights the effectiveness of frequency-based token weighting in improving the representation and prediction of rare items. The tail improvements are particularly pronounced on the \textit{Musical Instruments} and \textit{Yelp} datasets, which exhibit higher sparsity and stronger long-tailed distributions. 
Importantly, performance on head items is also improved across all datasets, indicating that the multi-target framework with curriculum learning helps the model to achieve balanced gains without sacrificing accuracy on popular items.

\begin{figure}[t]
    \centering
    \includegraphics[width=0.95\linewidth]{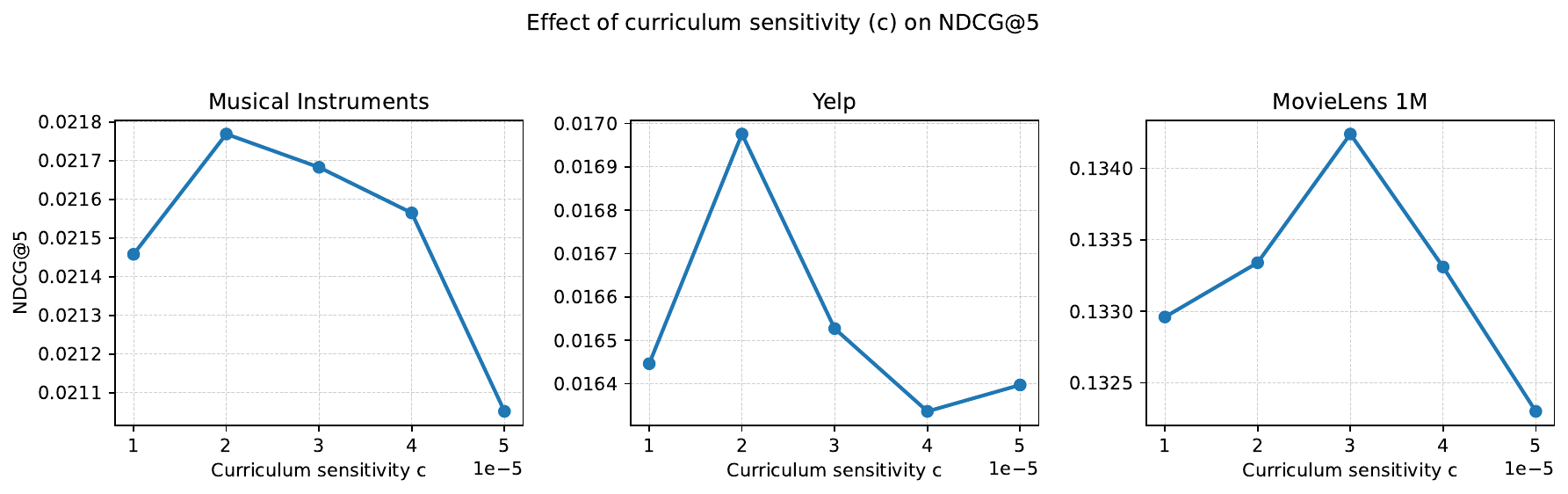}
    \caption{Performance of our method on different values of $c$, which controls how quickly the curriculum transitions during training.}
    \label{fig:c}
\end{figure}
\begin{figure}[t]
    \centering
    \includegraphics[width=0.95\linewidth]{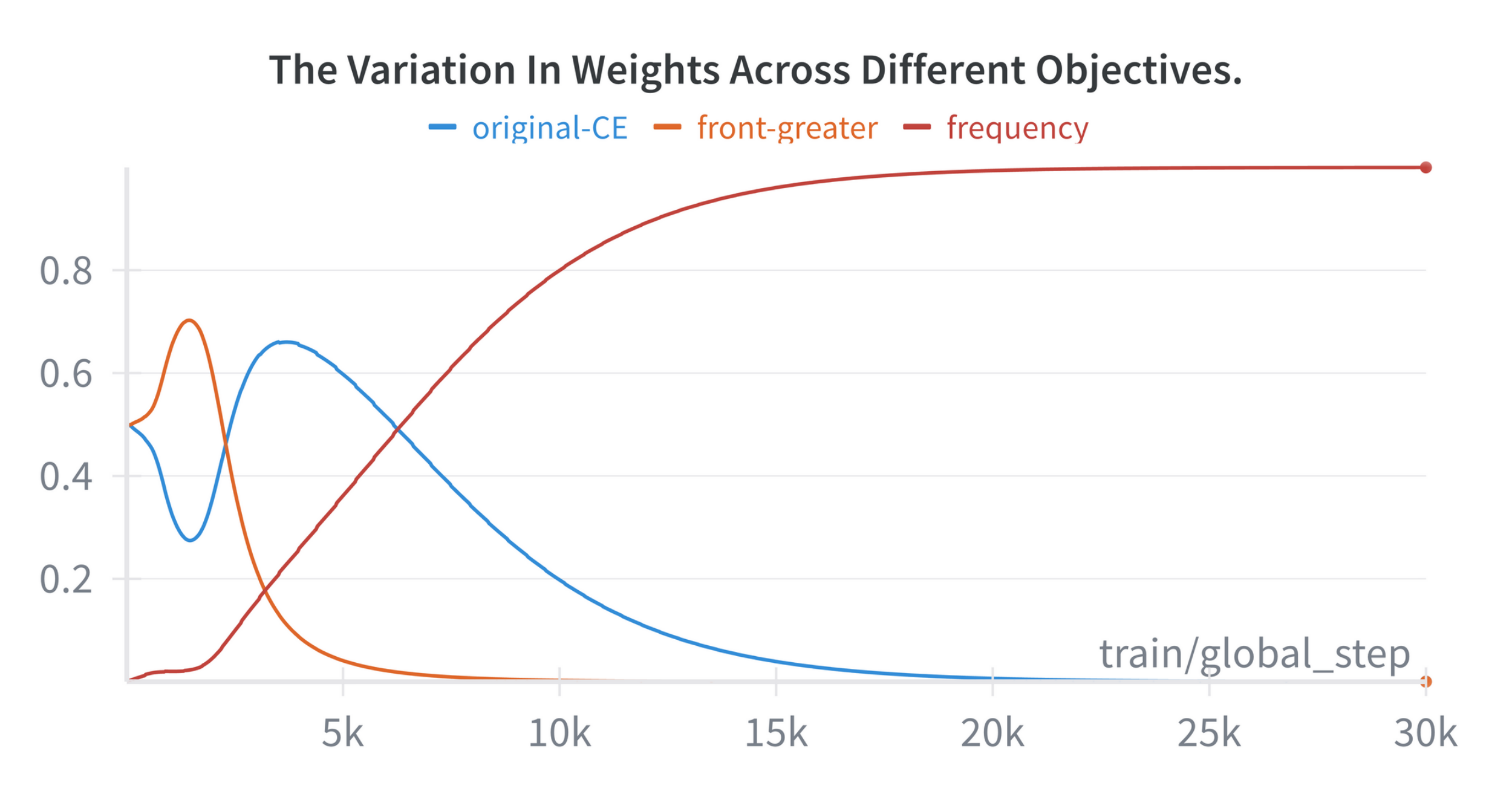}
    \caption{The variation in weights across different objectives on Yelp dataset with $c=2e-5$.}
    \label{fig:weight}
\end{figure}
\subsubsection{\textbf{Curriculum Progression Analysis (RQ6)}}
The hyperparameter $c$ controls the transition speed of the curriculum. Larger values induce a faster shift from Front-Greater Weighting and the original cross-entropy objective toward Frequency Weighting. We vary $c$ from $1\mathrm{e}{-5}$ to $5\mathrm{e}{-5}$ and report NDCG@5 in Figure~\ref{fig:c}.

Overall performance improves as $c$ increases from $1\mathrm{e}{-5}$, with peak results typically achieved at $2\mathrm{e}{-5}$ and $3\mathrm{e}{-5}$. When the curriculum progresses too slowly, the model overemphasizes early, coarse-grained tokens and lacks sufficient training time to learn from later, fine-grained tokens. In contrast, increasing $c$ beyond $3\mathrm{e}{-5}$ degrades performance, as transitioning too early to Frequency Weighting limits the warm-up provided by Front-Greater Weighting and standard cross-entropy. These results indicate that a moderate curriculum transition rate yields the best trade-off between training stability and fine-grained discrimination.

To illustrate how objectives are dynamically balanced in practice, in Figure~\ref{fig:weight}, we record the learned objective weights of a run on Yelp with $c=2e-5$. Initially, Front-Greater and standard cross-entropy are equally weighted, while Frequency Weighting is inactive. During early training (0–2.5k steps), Front-Greater rapidly dominates, emphasizing coarse semantic alignment. From 3k to 6k steps, the model gradually shifts toward standard cross-entropy, stabilizing full-sequence generation. After approximately 3k steps, the weight of Frequency Weighting rises sharply and eventually dominates (after around 14k steps), allowing the model to focus on rare-token discrimination. This progression aligns well with the intended coarse-to-fine learning dynamics and explains the strong empirical performance observed.
\section{Computational Analysis of Token Weighting}

We analyze the computational complexity of the proposed token-weighting strategies, namely \textbf{Frequency Token Weighting} and \textbf{Front-Greater Token Weighting}. Let $N$ denote the number of training instances, $L$ the token-ID length (i.e., number of tokens per semantic ID), $V = |\mathcal{V}|$ the vocabulary size, and $d$ the dimensionality of item embeddings. Importantly, both weighting schemes are computed \textbf{offline} as a preprocessing step and therefore introduce negligible overhead during model training.

\subsection{Offline Weight Computation.}
For Frequency Token Weighting, we compute the occurrence count $n_t$ for each token $t \in \mathcal{V}$ via a single pass over the dataset. Since each instance contains $L$ tokens, this step requires $\mathcal{O}(N L)$ time. The corresponding weights, derived from the effective number of samples, are then computed in $\mathcal{O}(V)$ time.

For Front-Greater Token Weighting, we estimate semantic conditional information gain by measuring the reduction in embedding dispersion across prefix-induced groups. By organizing all semantic ID sequences into a prefix tree, we can aggregate group-wise embedding statistics efficiently with a total cost of $\mathcal{O}(N L d)$. This avoids redundant regrouping operations and ensures scalability.

Since both weighting schemes are computed once prior to training, their computational cost does not affect training-time efficiency.

\subsection{Training-Time Complexity.}
During training, each token is associated with a precomputed scalar weight. The additional memory overhead is therefore $\mathcal{O}(V)$ for Frequency Weighting and $\mathcal{O}(L)$ for Front-Greater Weighting. Applying the weights during loss computation requires only constant-time lookup and element-wise multiplication. As a result, the overall computational complexity per training step remains unchanged compared to the original TIGER framework, which is dominated by the autoregressive softmax computation.

\section{Conclusion}

In this paper, we propose two complementary information-gain–based token-weighting strategies tailored to generative recommenders with semantic item identifiers. \textbf{Front-Greater Weighting} captures conditional semantic information gain by prioritizing early-position tokens that encode coarse, highly discriminative semantics and play a decisive role in autoregressive decoding. \textbf{Frequency Weighting} models marginal information gain under long-tailed token distributions, upweighting rare but informative tokens to counteract popularity bias. We further formulate training as a \textbf{multi-target learning framework} with \textbf{curriculum learning}, enabling stable optimization and effective integration with standard likelihood. Experiments on multiple datasets show consistent and significant improvements, confirming the effectiveness and generality of our approach. Building on this foundation, future work can extend to alternative item representations and incorporate additional objectives, fairness~\cite{itemfair, facter, up5}, diversity~\cite{personaldiversify, intentdiversify}, and user-intent–aware optimization~\cite{llm4isr}, within the same multi-target framework.
\section{GenAI Usage Disclosure}
Generative AI tools were used for language polishing, improving clarity and readability of the manuscript, and providing lightweight implementation assistance such as debugging suggestions and code refinement. All core research ideas, methodological designs, theoretical derivations, experiments, and final writing decisions were developed and verified by the authors.
\bibliographystyle{ACM-Reference-Format}
\bibliography{myref_clean}

\end{document}